\documentclass[10pt,journal]{IEEEtran}
\usepackage{epsfig}
\usepackage{amsmath}
\usepackage{amstext}
\usepackage{amsfonts}
\usepackage{amssymb}
\usepackage{eucal}
\usepackage{graphicx}
\usepackage{verbatim}
\usepackage{stmaryrd}
\usepackage{pstricks}
\usepackage{pst-node}
\usepackage{psfrag}
\usepackage{booktabs}
\usepackage{tikz}
\newcommand{\pw}[1]{\psframebox[linewidth=0.4pt,framearc=.3]{#1}}

\DeclareMathOperator{\diag}{diag}

\newcommand{\tr}[1]{{{\rm tr}_{#1}}}
\DeclareMathOperator{\EE}{E}
\newtheorem{definition}{Definition}

\begin{document}
\bibliographystyle{IEEEtran}

\title{Cognitive OFDM network sensing:\\ a free probability approach}

\author{\authorblockN{Romain Couillet\\}
\authorblockA{ST-NXP Wireless - Supelec\\
505 Route des Lucioles\\
06560 Sophia Antipolis, France\\
Email: romain.couillet@nxp.com} \\ \vspace{.3cm}
\and
\authorblockN{M{\'e}rouane~Debbah\\}
\authorblockA{Alcatel-Lucent Chair, Supelec\\
Plateau de Moulon, 3 rue Joliot-Curie \\
91192 Gif sur Yvette, France\\
Email: merouane.debbah@supelec.fr}
}

\maketitle

\begin{abstract}
  In this paper, a practical power detection scheme for OFDM terminals, based on recent free probability tools, is proposed. The objective is for the receiving terminal to determine the transmission power and the number of the surrounding base stations in the network. However, the system dimensions of the network model turn energy detection into an under-determined problem. The focus of this paper is then twofold: (i) discuss the maximum amount of information that an OFDM terminal can gather from the surrounding base stations in the network, (ii) propose a practical solution for blind cell detection using the free deconvolution tool. The efficiency of this solution is measured through simulations, which show better performance than the classical power detection methods.
\end{abstract}

\section{Introduction}
The ever increasing demand of high data rate has pushed system designers to exploit the wireless channel medium to the smallest granularity. In this respect, the orthogonal frequency division multiplexing (OFDM) modulation has been chosen as the next common standard in most wireless communication systems, e.g. WiMax \cite{WiMax}, 3GPP-LTE \cite{LTE}. OFDM converts a frequency selective fading channel into a set of flat fading channels \cite{BIN90}, therefore providing a high flexibility in terms of power and rate allocation. Future wireless networks therefore tend to be based on highly loaded OFDM cells. However, in multiple cell environments, inter-cell interference is still the bottleneck factor which considerably reduces the network-wide capacity. Cooperation between base stations are envisioned to reach the capacity performance of the so-called {\it broadcast channel} \cite{CAI03}, but many problems (essentially of power allocation and user scheduling) prevent those solutions to appear soon in practical standards. Therefore, it is essential for mobile terminals to be able to determine which neighboring cell provides the best quality of service, so that the terminal quickly hands over this best performance base station. Classically, only scarce and narrow-band pilot sequences allow the terminals to estimate the transmission power of the main surrounding base stations, e.g. in 3GPP-LTE, two sequences of the $0.7~{\rm MHz}$ band are available every $5~{\rm ms}$. Those synchronization sequences are usually affected by fast channel fading and overlap data from other base stations; as a consequence numerous occurrences of those pilots need be accumulated to achieve a satisfying estimation of the base stations transmission power.

The classical alternative to the pilot-aided (also referred to as {\it data-aided}) power detection is to perform a blind estimation from the incoming interfering signals. This raises the fundamental cognitive radio question \cite{MIT99}, \cite{HAY05}, which will be an important topic of the present work: ``how much information can a cognitive receiver recover from the incoming signals?''. The response to this question answers two classical concerns of engineers and system designers: (i) is the additional information brought by blind detection worth the computational effort?, (ii) is some given blind detector solution far from providing all the accessible information?. It is clear in particular, from an information theoretic viewpoint, that the information received on the $N$ OFDM subcarriers must ideally not be filtered in order to provide as much information as possible on the problem at hand, i.e. any filtering process diminishes the available information in the Shannon's sense \cite{SHA48}. Therefore, if as many as $L$ consecutive OFDM symbols are received, the available information is contained in the received $N\times L$ matrix ${\bf Y}$, with $N$ typically large. As a consequence, since $L$ cannot be taken infinitely large, $N/L$ is non trivial. This leads to the study of large random matrices problems, which is currently a hot topic in the wireless communication community \cite{RMT}. This is in sharp contrast with classical power detection methods \cite{URK67}, \cite{KOS02} which are only asymptotically unbiased, i.e. these methods assume that one of the system dimensions is large with respect to the others and this condition is necessary to ensure the convergence of the underlying algorithms.

Our purpose is to retrieve relevant information on the base station transmission powers. It will be shown hereafter that, depending on the {\it a priori} knowledge of the receiver, the essential part of the power information is, in most practical situations, contained in the {\it eigenvalue distribution} of the matrix $\frac{1}{L}{\bf Y}{\bf Y}^{\sf H}$. This naturally leads to the consideration of recent research on random matrix theory (RMT) \cite{RMT} and more specifically on free deconvolution \cite{RYA06}. In particular, in \cite{RYA07}, a similar study of terminal power detection in code division multiple access (CDMA) networks is derived from these tools. However, the model in \cite{RYA07} only considers flat fading channels and dodges the difficulty of multi-path channels; moreover the structure of the CDMA encoding matrix allows to easily recover the transmitted signal variances, which is not the case of multi-cell OFDM in which multiple streams overlap with no dedicated code to separate them. 

We propose here first to discuss the optimal amount of information which the receiver can extract from the incoming data to blindly retrieve the values of the powers transmitted by all surrounding base stations, when the receiver's prior state of knowledge about the environment is very limited. From this analysis, it will be observed that the general problem is not very tractable both in terms of mathematical derivation and therefore in terms of practical implementation. Secondly, we propose a suboptimal but implementable approach to solve the cell detection problem, based on the free probability framework, for which we derive a novel signal detection algorithm for OFDM.

The remainder of the paper is structured as follows: In Section \ref{sec:model}, we introduce the model of the multiple cell OFDM network. In Section \ref{sec:info}, we discuss the amount of information about transmitted powers which can be collected by the terminal from the received matrix ${\bf Y}$. The observation that all the necessary information is contained in the eigenvalues of $\frac{1}{L}{\bf Y}{\bf Y}^{\sf H}$ leads to Section \ref{sec:class}, in which we evaluate the classical energy detection methods, which assume $L/N\to \infty$. In Section \ref{sec:RMT}, we introduce some basic concepts of random matrix theory \cite{RMT}, which are needed to the understanding of the subsequent sections. In Section \ref{sec:algo}, we provide a novel algorithm to detect the cell transmission powers. Simulation results are then provided in Section \ref{sec:results}. A discussion on the gains and limitations of this novel method is carried out in Section \ref{sec:discussion}. Finally, in Section \ref{sec:conclusion} we draw our conclusions.

\textit{Notations:~} In the following, boldface lower case symbols represent vectors, capital boldface characters denote matrices (${\bf I}_N$ is the size-$N$ identity matrix). The spaces $\mathcal M(\mathcal A,i,j)$ and $\mathcal M(\mathcal A,i)$ are the sets of $i\times j$ and $i\times i$ matrices over the algebra $\mathcal A$, respectively. The transpose and Hermitian transpose operators are denoted $(\cdot)^{\sf T}$ and $(\cdot)^{\sf H}$, respectively. The operator $\diag({\bf x})$ turns the vector $\bf x$ into a diagonal matrix. 
The function ${\bf 1}_{(\cdot)}$ denotes the indicator function. 

\section{Downlink Model}
\label{sec:model}

\begin{figure}[t]
\psfrag{BS1}{\hspace{-.4cm}${\bf s}^{(l)}_1$}
\psfrag{BS2}{\hspace{-.35cm}${\bf s}^{(l)}_2$}
\psfrag{BS3}{\hspace{0cm}${\bf s}^{(l)}_3$}
\psfrag{BSk}{\hspace{.5cm}${\bf s}^{(l)}_M$}
\centerline{\includegraphics[width=8cm]{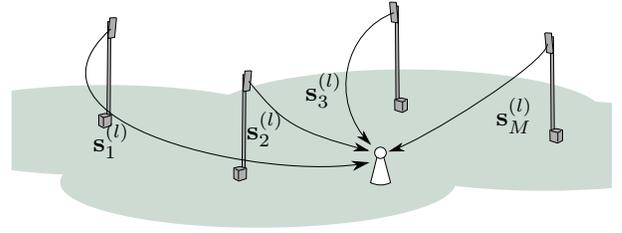}}
\caption{System Model}
\label{fig:scenario}
\end{figure}

Consider a network of $N_B$ base stations and one terminal equipped with a single receiving antenna. This scenario is depicted in Figure \ref{fig:scenario}. Assume moreover that the receiver is already connected to one serving base station, which we purposely exclude from the set of the $N_B$ surrounding base stations since we already know all information about it. The network is assumed to be synchronized in time and frequency and to use OFDM modulation with a size $N$ discrete Fourier transform (DFT). Denote then $M$ the maximum number of base stations the receiver expects to detect. Ideally $N_B\leq M$. In the following, for simplification, we assume that this condition is always fulfilled and we will only consider the parameter $M$, considering then a network of $M$ base stations, of which some could be of null power. The link between the terminal and the base station $k$ is modelled as a fast-fading complex frequency domain channel ${\bf h}_k=[h_{k1}\ldots h_{kN}]^{\sf T}\in \mathbb C^N$, coupled to a slow-fading path loss $L_k=1/P_k$ with $P_k$ the mean received power originating from base station $k$. The terminal also receives additive white Gaussian noise $\sigma{\bf n}\in \mathbb C^N$ with entries of variance $\sigma^2$. The base station $k$ sends at time $l$ the frequency-domain OFDM symbol ${\bf s}^{(l)}_k=[s_{1k}^{(l)},\ldots,s_{Nk}^{(l)}]^{\sf T}$. Therefore, the received signal vector ${\bf y}^{(l)}=[y_1^{(l)},\ldots,y_N^{(l)}]^{\sf T}$ at time $l$ reads

\begin{equation}
{\bf y}^{(l)}   = \sum_{k=0}^{M-1}P_k^{\frac{1}{2}}{\bf D}_k{\bf s}^{(l)}_k + \sigma{\bf n}^{(l)}
\end{equation}
with ${\bf D}_k=\diag({\bf h}_k)$.

This summation over the $M$ cells can be rewritten
\begin{eqnarray}
{\bf y}^{(l)} &= {\bf H}{\bf P}^{\frac{1}{2}}{\boldsymbol \theta}^{(l)}+\sigma{\bf n}^{(l)}
\end{eqnarray}
with ${\boldsymbol \theta}^{(l)}\in \mathbb C^{MN}$ the concatenated vector ${\boldsymbol \theta}^{(l)} = [{\bf s}^{(l)}_1,\ldots,{\bf s}^{(l)}_M]^{\sf T}$, 
${\bf H}\in \mathcal M(\mathbb C,N,MN)$ the concatenated matrix of the ${\bf D}_k$, $k\in \{1,\ldots,M\}$
\begin{equation}
{\bf H}=
\begin{bmatrix}
h_{11} & \cdots & 0      & \cdots & h_{M1} & \cdots & 0       \\
\vdots & \ddots & \vdots & \cdots & \vdots & \ddots & \vdots  \\
0      & \cdots & h_{2N} & \cdots & 0      & \cdots & h_{MN}
\end{bmatrix}
\end{equation}
and ${\bf P}\in \mathcal M(\mathbb R,N)$ the diagonal matrix
\begin{eqnarray}
{\bf P} &= 
\begin{bmatrix} P_1 & 0 & \cdots & 0 \\ 0 & P_2 & \ddots & \vdots  \\ \vdots & \ddots & \ddots & 0 \\ 0 & \cdots & 0 & P_M \end{bmatrix} 
\otimes {\bf I}_N
\end{eqnarray}

Now assume that the $M$ channels have a coherence time of order (or more than) $L$ times the OFDM symbol duration. The $L$ samples ${\bf y}^{(l)}$, $l=1,\ldots, L$, can be concatenated into an $N \times L$ matrix $\bf Y=[{\bf y}^{(1)}~\cdots~{\bf y}^{(L)}]$ to lead to the more general matrix expression of the received signal ${\bf Y}$
\begin{equation}
\label{eq:prod}
{\bf Y} = {\bf H}{\bf P}^{\frac{1}{2}}{\bf \Theta} + \sigma{\bf N}
\end{equation}
where ${\bf \Theta} \in \mathcal M(\mathbb C,MN,L)$ and ${\bf N}\in \mathcal M(\mathbb C,N,L)$ are the concatenation matrices of the $L$ vectors ${\boldsymbol \theta}^{(l)}$ and ${\bf n}^{(l)}$, respectively.

\section{Problem statement}
\label{sec:info}
The power detection problem consists in the present situation to retrieve the entries of the matrix $\bf P$ from the receive matrix $\bf Y$. The prior information at the receiver is not sufficient to find $\bf P$ in a straightforward manner. Indeed, the channel matrix $\bf H$ and the transmitted signal $\bf \Theta$ are unknown and, worse, even their stochastic distribution are usually unknown; in particular, recent flexible multiple access OFDM standards adapt their transmission rates to the channel quality so that the terminal cannot {\it a priori} assume to receive either QPSK, $16$-QAM, $64$-QAM or any other type of modulation. The {\it a priori} knowledge $I$ at the terminal is therefore limited to: (i) the approximated background covariance $\EE[{\bf n}{\bf n}^{\sf H}]=\sigma^2{\bf I}$, (ii) the fast-fading channel power ${\rm E}[{\bf h}_k^{\sf H}{\bf h}_k]=1$, (iii) the channel delay spread known to be lesser than the cyclic prefix length, (iv) the transmitted signal covariance ${\rm E}[{\boldsymbol \theta}_k^{\sf H}{\boldsymbol \theta}_k]={\bf I}_N$.

From this amount of prior information $I$, the most reliable channel model\footnote{by ``most reliable model'', we mean the model which satisfies the constraints imposed by $I$ and which is the most noncommittal regarding unknown system information.} is obtained from the maximum entropy principle \cite{JAY57}. The latter states that the transmitted data $\bf \Theta$ must be modelled as a Gaussian independent and identically distributed process (i.i.d.). As for the short-term channels ${\bf h}_k$, given a delay spread $\tau_d$ (counted in integer number of time-domain samples), the time-domain representation of ${\bf h}_k$ must be modelled as a Gaussian i.i.d. vector of length $\tau_d$; therefore, ${\bf h}_k$ is to be modelled as a Gaussian vector with covariance matrix the DFT of ${\bf I}_{\tau_d}$. Since little information about $\tau_d$ is initially known to the receiver, the channels must be modelled as the marginal distribution of those Gaussian processes with $\tau_d$ varying from $1$ to the cyclic prefix length.

Of course, this model might be very different from reality and might provide totally wrong results, as longly discussed in \cite{JAY03}. However, this is the best one can blindly infer on the transmission scheme from the available information. The objective now is to determine what is the probability $p({\bf P}|I)$ that a sequence of transmitted powers $\{P_1,\ldots,P_M\}$ fits the previous model knowing $I$. From those probabilities, computed for all vectors in $(\mathbb R^+)^M$, an estimate $\hat{\bf P}$ of $\bf P$ can be designed which minimizes some error measure, e.g. $\hat{\bf P}=\EE[{\bf P}|{\bf Y},I]$ would be the minimum mean square error estimate of $\bf P$.

The probability $p({\bf P}|{\bf Y},I)$ assigned to the information $({\bf P}|{\bf Y},I)$ can be written, thanks to Bayes' rule
\begin{align}
  p({\bf P}|{\bf Y},I) &= p({\bf Y}|{\bf P},I)\cdot \frac{p({\bf P}|I)}{p({\bf Y}|I)}
\end{align}
in which $({\bf P}|I)$ is the {\it a priori} knowledge about ${\bf P}$. It is classically assigned a uniform distribution over some subspace $[0,P_{\sf max}]^M$ for a maximum receive power $P_{\sf max}$. As for $p({\bf Y}|{\bf P},I)$, it can be expanded as
\begin{align}
  \label{eq:proba}
  p({\bf Y}|{\bf P},I) &= \int_{{\bf \Theta},{\bf H}} p({\bf Y}|{\bf P},{\bf H},{\bf \Theta},I) p({\bf H}|I) p({\bf \Theta}|I) {\rm d}{\bf H}{\rm d}{\bf \Theta}
\end{align}
in which all integrands are known from the maximum entropy model aforementioned,
\begin{itemize}
  \item $p({\bf \Theta}|I)$ is standard multivariate i.i.d. Gaussian.
  \item the compound channel $\bf H$ is assigned a distribution
    \begin{equation}
      \label{eq:channeldist}
      p({\bf H}|I)=\sum_{k=1}^{\tau_{\sf max}}p({\bf H}|\tau_d,I)\cdot p(\tau_d|I)
    \end{equation}
    with $\tau_d$ the channel delay spread, $\tau_{\sf max}$ the cyclic prefix length, $p(\tau_d|I)=1/\tau_{\sf max}$ and $p({\bf H}|\tau_d,I)$ with standard i.i.d. Gaussian diagonals.
\end{itemize}

However, the explicit computation of \eqref{eq:proba} is very involved and requires advanced tools from random matrix theory. A similar calculus was performed by the authors for the simpler single-cell MIMO energy detector \cite{COU08}. In the latter it was shown that, surprisingly, the standard i.i.d. Gaussian model assigned to the main system parameters makes the energy detection depend only on the eigenvalue distribution of the receive matrix ${\bf Y}$. The multi-cell detection problem at hand is very similar in configuration, apart for the channel marginalization of equation \eqref{eq:channeldist} which is not i.i.d. Gaussian. Since we cannot provide an optimal information theoretical solution to our problem and since both aforementioned problems are very similar, it seems relevant to concentrate on the close-to-optimal random matrix theoretical approach. 

Some important information can nonetheless be already deduced from the integral form of equation \eqref{eq:proba}. If the transmission channels are extremely frequency flat, i.e. for all $k$, $h_{k1}\simeq h_{k2} \ldots \simeq h_{kN}$, then $\{{\bf H}{\bf P}^{\frac{1}{2}}{\bf \Theta}\}_{ij}=\sum_k\sqrt{P_k}h_k \theta_{kj}$. Therefore, even if the realizations of ${\bf N}$ and $\bf \Theta$ were perfectly known, one will have access at best to the variables $\sqrt{P_k}h_k$, $k=1,\ldots,M$, from which no reliable estimation of $P_k$ be drawn; in such a situation, the posterior probability $p({\bf P}|{\bf Y},I)$ is very broad and is maximized on a large continuous set of $P_1,\ldots,P_M$. On an information theoretical viewpoint, this means that the optimal inference on $\bf P$ given $\bf Y$ and $I$ cannot lead to any valuable information.
In the random matrix approach, the situation is even worse. If one knew perfectly the entries of ${\bf H}{\bf P}{\bf H}^{\sf H}$, then nothing at all can be said about $P_1,\ldots,P_M$. Indeed, $\{{\bf H}{\bf P}{\bf H}^{\sf H}\}_{ij}=\sum_k P_k |h_k|^2$ (see Section \ref{sec:algo} for details) and the only piece of information which one has about $P_1,\ldots,P_M$ is the sum $\sum_k P_k |h_k|^2$; the latter cannot lead to any estimate of ${\bf P}$ when $M>1$ and the problem cannot be solved.

This means that, given the limited prior information of the terminal, it is {\it impossible} to come up with a reliable estimate of $\bf P$ when the channels are frequency flat. In the remainder of this paper, we shall therefore consider that the OFDM channels are very frequency selective\footnote{note that this assumption ensures high efficiency of the network in terms of per-user outage capacity, which is very desirable in the current trend for packet-switched communications.}. In the following, we investigate the classical power detection techniques, which shall prove inefficient in this large non-trivial matrix problem.

\section{Classical Power Detection}
\label{sec:class}
Usual power detection considers the second order statistics of the received signals. In the scalar case, i.e. ${\bf y}^{(l)}$ reduces to a single value $y^{(l)}$, it was proved \cite{URK67} that the optimal detector with the aforementioned state of knowledge at the terminal consists in evaluating $\frac{1}{L}([y^{(1)},\ldots,y^{(L)}][y^{(1)},\ldots,y^{(L)}]^{\sf H})-\sigma^2$, with $L>>1$. We show in what follows that this classical scheme can be simply extended to our network situation but that it is very inefficient for small $L/N$ ratios. 

Assuming that $L/N$ is very large, the expression of the normalized Gram matrix associated to ${\bf Y}$ reads
\begin{eqnarray}
\frac{{\bf YY}^{\sf H}}{L} &{\underset{L\to\infty}{\longrightarrow}}& {\mathbb E}\left[\frac{1}{L} ({\bf H}{\bf P}^{\frac{1}{2}}{\bf \Theta} + \sigma{\bf N})({\bf H}{\bf P}^{\frac{1}{2}}{\bf \Theta} + \sigma{\bf N})^{\sf H}\right] \nonumber \\
&{\underset{L\to\infty}{\longrightarrow}}& {\bf HP}^{\frac{1}{2}}{\mathbb E}\left[\frac{{\bf \Theta}{\bf \Theta}^{\sf H}}{L} \right] {\bf P}^{\frac{1}{2}}{\bf H}^{\sf H} +\frac{{\bf HP}^{\frac{1}{2}}}{L}{\mathbb E}\left[{\bf \Theta N}^{\sf H} \right] \nonumber \\
&&+ \frac{1}{L}{\mathbb E}\left[{\bf N\Theta}^{\sf H} \right]{\bf P}^{\frac{1}{2}}{\bf H} + {\mathbb E}\left[\frac{1}{L}{\bf NN}^{\sf H}\right] \nonumber \\
&{\underset{L\to\infty}{\longrightarrow}}&  {\bf HPH}^{\sf H} + \sigma^2{\bf I}_N
\end{eqnarray}
the last line comes from the fact that, $N$ being finite, the $N\times N$ matrices $\frac{1}{L}{\bf NN}^{\sf H}$ and $\frac{1}{L}{\bf \Theta\Theta}^{\sf H}$ converge in distribution to an identity matrix, and the cross products to null matrices.

As a consequence, as will be detailed in Section \ref{sec:findPk}, one can estimate the values $P_k$, $k\in \{1,\ldots,M\}$ from the moments $\{(\frac{1}{L}{\bf YY}^{\sf H}-\sigma^2{\bf I}_N)^k\}$, $k\in \{1,\ldots,M\}$, when $L$ is large compared with $N$.

Our situation does not fall into this asymptotic $L/N\to \infty$ context. In present and future OFDM technologies, the number $N$ of available subcarriers is large, e.g. of order a thousand subcarriers, while $L$ is limited in our model to the channel coherence time or in general to the number of OFDM symbols the terminal is willing to memorize before treating information. Therefore, even if $N$ and $L$ are large, their ratio $N/L$ is not in general close to zero. The fundamental asymptotic assumption is therefore no longer satisfied. 

We show in Table \ref{tab:classical} that the non-trivial ratio $N/L$ impairs significantly the performance of the classical power detection. The latter is the result of a simulation in which we applied the algorithm that will be described in Section \ref{sec:findPk}, based on the sample moments of $\frac{1}{L}({\bf YY}^{\sf H}-\sigma^2{\bf I}_N)$ (instead of the sample moments of $\frac{1}{L}{\bf HPH}^{\sf H}$, whose corresponding results are shown between brackets). The typical situation considered in this example is a three-base station scenario of respective powers $P_1=4$, $P_2=2$ and $P_3=1$ and a noise level $\sigma^2=0.1$, i.e. ${\sf SNR}=10~{\rm dB}$, $N=256$ and $L$ is taken in a range from $256$ to $32,768$. It turns out indeed that $L$ needs to be large for this method to be satisfying. In this precise example, this compels $L/N$ to be of order $64$, which is not acceptable in our current system settings.

\begin{table}[t]
\centerline{
\begin{tabular}{rcr}
\toprule
\multicolumn{3}{c}{\raisebox{-.5ex}{$N=256$, ${\bf P}=\{P_1,P_2,P_3\}=\{4,2,1\}$}} \\ [1ex]
\toprule
\raisebox{-.7ex}{$L$} & \raisebox{-.7ex}{Estimated $\tilde{\bf P}$ [our algorithm]} & \raisebox{-.7ex}{$\Vert{\bf P}-\tilde{{\bf P}}\Vert_2$} \\ [1ex]
\midrule
$256$ & $\{7.93,1.62,-2.5\}$ [$\{4.28,1.35,1.35\}$] & $27.63$ \\
$512$ & $\{5.82,2.90,-1.7\}$ [$\{3.80,2.16,1.00\}$] & $11.42$  \\
$1024$ & $\{4.26,3.60,-0.8\}$ [$\{3.62,2.37,0.94\}$] & $5.87$  \\
$2048$ & $\{4.52,2.69,-0.2\}$ [$\{4.22,1.55,1.12\}$] & $1.77$ \\
$4096$ & $\{4.20,2.65,0.18\}$ [$\{4.09,2.06,0.78\}$] & $1.41$ \\
$8192$ & $\{4.10,2.28,0.58\}$ [$\{4.05,1.89,0.92\}$] & $0.27$ \\
$16384$ & $\{3.97,2.42,0.89\}$ [$\{3.95,2.24,0.99\}$] & $0.19$ \\
$32768$ & $\{4.07,1.95,0.99\}$ [$\{4.03,1.95,0.98\}$] & $0.01$ \\
\bottomrule
\addlinespace[2\aboverulesep]
\end{tabular}
}
\caption{Classical moment-based method}
\label{tab:classical}
\end{table}

Such problems involving large matrices with non-trivial $N/L$ ratios are at the heart of a recent field of research, known as random matrix theory (RMT), which is a particular case of the more general free probability theory introduced by Voiculescu \cite{VOI92}. In the subsequent section, we provide a quick introduction to important notions of RMT which are necessary to handle the rest of the multiple cell detection study.

\section{RMT and free deconvolution}
\label{sec:RMT}

\subsection{Random matrix theory}
\label{MRMT}
\begin{definition}
A random matrix is a multi-variate random variable ${\bf X}=\{X_{11},X_{12},\ldots,X_{MN}\}$ for a given $(M,N)\in \mathbb N^2$ couple. As such, $\bf X$ is a matrix whose entries $X_{ij}\in \mathbb C^{M\times N}$ are ruled by a joint probability distribution $p(X_{11},X_{12},\ldots,X_{MN})$.
\end{definition}

Free probability is the study of random variables in non-commutative algebras, i.e. algebras in which the product operation is non-commutative. The algebra of large Hermitian random matrices is a particular case of those non-commutative algebras. In the following, we shall qualify \textit{free} any notion attached to the free probability (or RMT) framework while we shall qualify \textit{classical} any notion attached to the classical probability framework of commutative algebras.

Similarly to the classical theory of probability, a free \textit{expectation} functional $\phi$ can be defined. For a given Hermitian random matrix $\bf X$, the free expectation reads
\begin{align}
\phi\left({\bf X}\right) &= \lim_{N\rightarrow \infty} \EE\left[\tr{N}{\bf X}\right]	
\end{align}
and we can similarly define free \textit{moments} $m_k,k\in \mathbb N$ of a random matrix. Those are
\begin{equation}
m_k= \phi\left( {\bf X}^k \right)
\end{equation}

Thanks to the trace properties, note that the free moments are strongly linked to the eigenvalues $\lambda_i$, $i\in \{1,N\}$ of $\bf X$, since $m_k$ can also be written
\begin{equation}
\label{eq:freemomentsf}
m_k = \lim_{N\rightarrow \infty} \frac{1}{N}\sum_{i=1}^N \lambda_i^k
\end{equation}

Indeed, denote ${\bf X}={\bf Q_{\bf \Lambda}}{\bf \Lambda}{\bf Q_{\bf \Lambda}}^{\sf H}$ with ${\bf \Lambda}=\diag(\{\lambda_1,\ldots,\lambda_N\})$ and ${\bf Q_{\bf \Lambda}}$ unitary, ${\bf X}^k={\bf Q_{\bf \Lambda}}{\bf \Lambda}^k{\bf Q_{\bf \Lambda}}^{\sf H}$. Taking the trace of ${\bf X}^k$ leads to equation \eqref{eq:freemomentsf}.

The asymptotic ($N,L\to \infty$ with $N/L$ constant) marginal distribution of the eigenvalues of $\bf X$ is called the \textit{empirical distribution} of $\bf X$ and will be denoted $\mu_{\bf X}$. Its associated cumulative distribution function $F_{\bf X}$ reads \cite{RMT}
\begin{equation}
  F_{\bf X}(\lambda)=\lim_{n\to \infty}\frac{1}{N}\sum_{i=1}^N {\bf 1}_{(\lambda_i\leq \lambda)}
\end{equation}

Therefore the free moments are directly linked to the empirical distribution of the matrix $\bf X$,
\begin{equation}
  m_k = \lim_{N\rightarrow \infty} \int_{\mathbb R^+} \lambda^k \mu_{\bf X}(\lambda){\rm d}\lambda
\end{equation}
which is the classical definition of moments associated to the distribution $\mu_{\bf X}$.

Interestingly, for most {\it usual}\footnote{by usual, we qualify matrices found in common wireless communication problems.} random matrices $\bf A$ of large dimensions $N,L\to \infty$ with $N/L=c$ constant, the eigenvalue density of ${\bf X}=\frac{1}{L}{\bf A}{\bf A}^{\sf H}$ converges to a definite distribution. For instance, in our current problem, since the input signals $\bf \Theta$ and noise $\bf N$ are modelled as standard i.i.d. Gaussian, we are interested in the so-called Wishart matrices that we define hereafter
\begin{definition}
  An $N\times N$ random matrix ${\bf X}={\bf AA}^{\sf H}$, with $\bf A$ a random $N\times L$ matrix whose columns are zero mean Gaussian vectors with covariance matrix $\Sigma$, is called a generalized Wishart matrix of $L$ degrees of freedom. This is denoted
\begin{equation}
\label{eq:wishart}
{\bf X} \sim \mathcal W_N(L,\Sigma)
\end{equation}
\end{definition}

\begin{figure}
  \begin{center}
  \includegraphics[width=8cm]{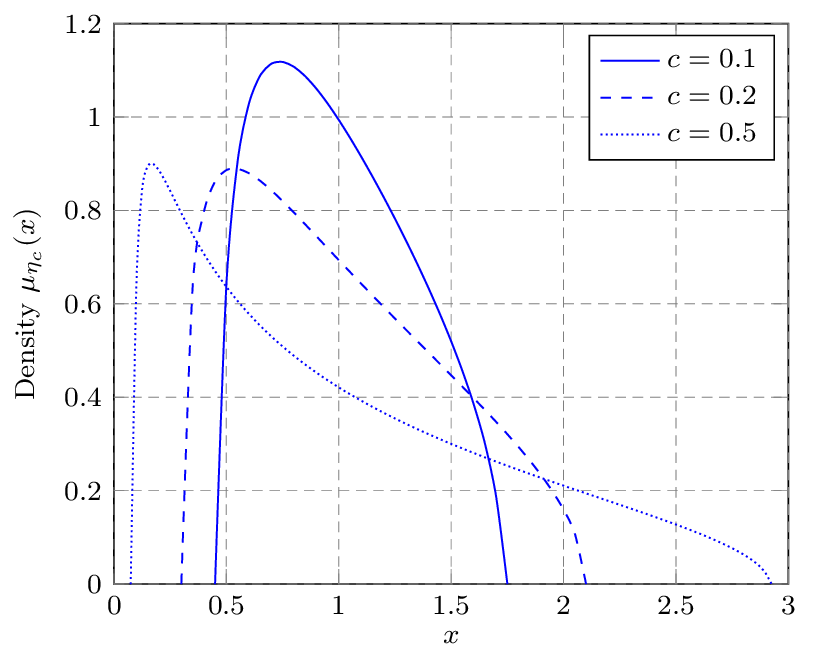}
  \caption{Marchenko-Pastur law $\mu_{\eta_c}$}
\label{fig:MP}
\end{center}
\end{figure}

Wishart matrices $W_N(L,{\bf I}_N)$ are known to have an eigenvalue distribution which converges, when $(N,L)$ grows to infinity with a constant ratio $c=N/L$, towards the Marchenko-Pastur law $\mu_{\eta_c}$ \cite{RMT}. The Marchenko-Pastur law is defined by
\begin{equation}
\mu_{\eta_c} = \left( 1- \frac{1}{c} \right)^+\delta(x)+\frac{\sqrt{(x-a)^+(b-x)^+}}{2\pi cx}
\end{equation}
with $(a,b)=\left((1-\sqrt{c})^2,(1+\sqrt{c})^2\right)$. In Figure \ref{fig:MP} we provide the distribution of $\mu_{\eta_c}$ for different values of $c$.

Note that when $c$ tends to $0$, i.e. when $L/N\to \infty$, the Marchenko-Pastur law converges to a single Dirac in $1$ and we recover the classical law of large numbers. 

Equivalently to classical probability theory, many results of free probability involve the distribution of sum, difference, product and inverse of random matrices. The characteristic function of a distribution, used to derive the distribution of the sum of \textit{independent} commutative random variables, has a free counterpart called the $R$-Transform. The Mellin transform, used to derive the product of \textit{independent} commutative random variables, also has a free counterpart, known as the $S$-Transform\footnote{note that \textit{independence} in the classical probability sense is not enough in free probability to derive the empirical distribution of the sum, difference, product and inverse product of two random matrices. This \textit{independence} notion is extended to the \textit{asymptotic freeness} concept \cite{RMT} which is not a technical issue in our study.}.

Given two large random Hermitian matrices $\bf A$ and $\bf B$, whose sum is ${\bf C}={\bf A}+{\bf B}$ and whose product is ${\bf D}={\bf AB}$, one can then derive the empirical distributions of $\bf C$ and $\bf D$ from the empirical distributions of $\bf A$ and $\bf B$, which we denote
\begin{align}
\label{eq:freeadd}
\mu_{\bf C}&=\mu_{\bf A}\boxplus \mu_{\bf B} \\
\label{eq:freetime}
\mu_{\bf D}&=\mu_{\bf A}\boxtimes \mu_{\bf B}
\end{align}
Equation \eqref{eq:freeadd} is called additive free convolution and equation \eqref{eq:freetime} is called multiplicative free convolution.
Similarly, given only the distributions of $\bf B$, $\bf C$ and $\bf D$, one can recover the distribution of $\bf A$
\begin{align}
\label{eq:freeminus}
\mu_{\bf A}&=\mu_{\bf C}\boxminus \mu_{\bf B} \\
\label{eq:freedivide}
\mu_{\bf A}&=\mu_{\bf D}\boxbslash \mu_{\bf B}
\end{align}
in which equation \eqref{eq:freeminus} is called additive free deconvolution and equation \eqref{eq:freedivide} is called multiplicative free deconvolution.

\subsection{Free deconvolution for information plus noise model}
\label{sec:FreeDeconv}
Our interest is to treat a particular communication model, known as the \textit{information plus noise model},

\begin{definition}
Given two $N\times L$ ($N/L\to c$) large random matrices ${\bf R}$ (standing for an informative signal) and $\bf X$ (standing for a noise additive signal) and a scalar $\sigma$ (the standard deviation of the noise process), the model given by
\begin{equation}
{\bf W}=\frac{1}{L}({\bf R}+\sigma {\bf X})({\bf R}+\sigma {\bf X})^{\sf H}
\end{equation}
is called the \textit{information plus noise} model.
We shall therefore call $\bf W$ an \textit{information plus noise} matrix.
\end{definition}

It has been recently shown \cite{RYA07} that the empirical distribution of $\frac{1}{L}{\bf RR}^{\sf H}$ in the previous definition can be recovered from the empirical distributions of the matrices ${\bf W}$ and ${\bf XX}^{\sf H}$ when $\bf X$ is Gaussian with i.i.d. entries of zero mean and variance $1/L$. This is given by
\begin{equation}
\label{eq:freedeconv}
\mu_{\frac{1}{L}{\bf RR}^{\sf H}}=\left( \left( \mu_{\bf W}\boxbslash \mu_{\eta_c}\right) \boxminus \delta_{\sigma^2} \right) \boxtimes \mu_{\eta_c}
\end{equation}
For more details about the demonstration of formula \eqref{eq:freedeconv}, refer to \cite{RYA06}.

Thanks to this RMT framework and the free convolution tools, we can now address our multiple cell detection problem.

\section{Application of free deconvolution to multiple cell detection}
\label{sec:algo}
\subsection{Signal and noise deconvolution}
In the model \eqref{eq:prod}, the $N\times N$ matrix $\frac{1}{L}{\bf Y}{\bf Y}^{\sf H}$ is an \textit{information plus noise} matrix with $\bf N$ a Gaussian random matrix. Therefore, according to equation \eqref{eq:freedeconv}, when $N$, $L$ are sufficiently large, one can derive the empirical distribution $\mu_{\frac{1}{L}{{\bf H}{\bf P}^{\frac{1}{2}}}{\bf \Theta\Theta}^{\sf H}{\bf P}^{\frac{1}{2}}{\bf H}^{\sf H}}$ from the empirical distribution $\mu_{\frac{1}{L}{\bf Y}{\bf Y}^{\sf H}}$ as follows
\begin{equation}
\label{eq:dec}
\mu_{\frac{1}{L}{{\bf H}{\bf P}^{\frac{1}{2}}}{\bf \Theta\Theta}^{\sf H}{\bf P}^{\frac{1}{2}}{\bf H}^{\sf H}} = \left( (\mu_{\frac{1}{L}{\bf Y}{\bf Y}^{\sf H}} \boxbslash \mu_{\eta_c}) \boxminus \delta_{\sigma^2} \right) \boxtimes \mu_{\eta_c}
\end{equation}
where $c=N/L$, for $\bf N$ is of size $N\times L$.

Also, the matrix ${\bf \Theta}$ in equation \eqref{eq:prod} is modelled as standard i.i.d. Gaussian. Therefore $\frac{1}{L}{\bf P}^{\frac{1}{2}}{\bf H}^{\sf H}{\bf H}{\bf P}^{\frac{1}{2}}{\bf \Theta\Theta}^{\sf H}$ is a \textit{generalized Wishart matrix} with covariance ${\bf P}^{\frac{1}{2}}{\bf H}^{\sf H}{\bf H}{\bf P}^{\frac{1}{2}}$. Analogously to \eqref{eq:wishart}, this can be written
\begin{align}
\frac{1}{L}{\bf P}^{\frac{1}{2}}{\bf H}^{\sf H}{\bf H}{\bf P}^{\frac{1}{2}}{\bf \Theta\Theta}^{\sf H} &\sim \mathcal W_N(L,{\bf P}^{\frac{1}{2}}{\bf H}^{\sf H}{\bf H}{\bf P}^{\frac{1}{2}})
\end{align}

Then, the empirical distribution of the covariance matrix ${\bf P}^{\frac{1}{2}}{\bf H}^{\sf H}{\bf H}{\bf P}^{\frac{1}{2}}$ of the Wishart matrix $\frac{1}{L}{\bf P}^{\frac{1}{2}}{\bf H}^{\sf H}{\bf H}{\bf P}^{\frac{1}{2}}{\bf \Theta\Theta}^{\sf H}$ can be recovered from the empirical distribution $\mu_{\frac{1}{L}{\bf P}^{\frac{1}{2}}{\bf H}^{\sf H}{\bf H}{\bf P}^{\frac{1}{2}}{\bf \Theta\Theta}^{\sf H}}$ when the couple $(N,L)$ tends to infinity with a constant ratio $c'=MN/L$ ($M$ is constant). Consequently, similarly to (\ref{eq:freedivide}),
\begin{equation}
\label{eq:lastdec}
\mu_{{\bf P}^{\frac{1}{2}}{\bf H}^{\sf H}{\bf H}{\bf P}^{\frac{1}{2}}} = \mu_{\frac{1}{L}{{\bf P}^{\frac{1}{2}}{\bf H}^{\sf H}{\bf H}{\bf P}^{\frac{1}{2}}}{\bf \Theta\Theta}^{\sf H}} \boxbslash \mu_{\eta_{c'}} 
\end{equation}

The left side of equation \eqref{eq:dec} is slightly different from the desired expression in the right side of equation \eqref{eq:lastdec}. Still, thanks to the trace commutativity property, we have the link \cite{RMT}
\begin{equation}
\label{eq:zeros}
\mu_{\frac{1}{L}{{\bf P}^{\frac{1}{2}}{\bf H}^{\sf H}{\bf H}{\bf P}^{\frac{1}{2}}}{\bf \Theta\Theta}^{\sf H}} = \frac{1}{M} \mu_{\frac{1}{L}{{\bf H}{\bf P}^{\frac{1}{2}}}{\bf \Theta\Theta}^{\sf H}{\bf P}^{\frac{1}{2}}{\bf H}^{\sf H}} + \left( 1-\frac{1}{M}\right)\delta_0
\end{equation}

This relation is due to the fact that the positive eigenvalues of ${\frac{1}{L}{{\bf H}{\bf P}^{\frac{1}{2}}}{\bf \Theta\Theta}^{\sf H}{\bf P}^{\frac{1}{2}}{\bf H}^{\sf H}}$ are the same as those of ${\frac{1}{L}{{\bf P}^{\frac{1}{2}}{\bf H}^{\sf H}{\bf H}{\bf P}^{\frac{1}{2}}}{\bf \Theta\Theta}^{\sf H}}$ (since their traces are identical and then all their moments match). But the rank of both matrices differ and then a certain amount of null eigenvalues must be introduced. Here, ${\frac{1}{L}{{\bf H}{\bf P}^{\frac{1}{2}}}{\bf \Theta\Theta}^{\sf H}{\bf P}^{\frac{1}{2}}{\bf H}^{\sf H}}$ is of full rank ($N$) while ${\frac{1}{L}{{\bf P}^{\frac{1}{2}}{\bf H}^{\sf H}{\bf H}{\bf P}^{\frac{1}{2}}}{\bf \Theta\Theta}^{\sf H}}$ is only of rank $N$ for a matrix size $MN$, hence the $M$ factor in equation \eqref{eq:zeros}.

Finally, we similarly connect the left side of equation \eqref{eq:lastdec} to $\mu_{{\bf H}{\bf P}{\bf H}^{\sf H}}$ through
\begin{equation}
\mu_{{\bf P}^{\frac{1}{2}}{\bf H}^{\sf H}{\bf H}{\bf P}^{\frac{1}{2}}} = \frac{1}{M} \mu_{{\bf H}{\bf P}{\bf H}^{\sf H}} + \left( 1-\frac{1}{M}\right)\delta_0
\end{equation}

The empirical distribution of ${{\bf H}{\bf P}{\bf H}^{\sf H}}$ was then derived from the empirical distribution of $\frac{1}{L}{\bf YY}^{\sf H}$. As a consequence, the free moments $d_k=\EE[\tr{N}({\bf H}{\bf P}{\bf H}^{\sf H})^k]$ can be retrieved from the free moments $m_k=\EE[\tr{N}(\frac{1}{L}{\bf Y}{\bf Y}^{\sf H})^k]$. Surprisingly, it is shown in \cite{RYA06} that for all aforementioned free convolution operations, the set of the first $k$ moments of the [de]convolved distributions can be recovered from the set of the $k$ first moments of the operands. This substantially reduces the computational effort, as is described in the following.

Let us work in detail all the steps to derive the moments $d_k$ from the moments $m_k$. 
\begin{enumerate}
  \item first, the noise contribution to the signal $\bf Y$ is deconvolved thanks to formula \eqref{eq:dec}. The multiplicative convolution (resp. deconvolution) $\mu_{({\sf out})}$ of a distribution $\mu_{({\sf in})}$ and the Marchenko-Pastur law $\mu_{\eta_c}$ can be computed from all the moments $m^{({\sf in})}_k$. It is shown in \cite{RYA07} that $c\cdot m^{({\sf out})}_k$ can be computed from the \textit{moments/cumulants} transform (resp. \textit{cumulants/moments} transform) \cite{RMT} of the coefficients $c\cdot m^{({\sf in})}_k$, with $c=N/L$. 
As for additive convolution (resp. deconvolution) $\mu_{({\sf add})}$ of two distributions $\mu_{({\sf a})}$ and $\mu_{({\sf b})}$, the free cumulants of $\mu_{({\sf add})}$ are the sum (resp. difference) of the cumulants of $\mu_{({\sf a})}$ and $\mu_{({\sf b})}$. 
From \eqref{eq:dec}, the moments $m_k'$ of $\mu_{\frac{1}{L}{\bf HP}^{\frac{1}{2}}{\bf \Theta\Theta}^{\sf H}{\bf P}^{\frac{1}{2}}{\bf H}^{\sf H}}$ are obtained from the moments $m_k$ of $\mu_{\frac{1}{L}{\bf YY}^{\sf H}}$; therefore, in mathematical terms, this reads
\begin{equation}
(m_1',\ldots,m_M')=\mathcal S_1(m_1,\ldots,m_M,\sigma^2)
\end{equation}
with
\begin{align}
\mathcal S_1&=\frac{1}{c}\mathcal M\left[ c\cdot \mathcal M\left( \mathcal C\left\{ \frac{1}{c}\mathcal C\left( cm_1,\ldots, cm_M \right) \right\} \right. \right. \nonumber \\ 
& \left. \left. -\mathcal C\left(\sigma^2,\ldots,\sigma^{2M} \right) \right) \right]
\end{align}
where the functions $\mathcal M(\cdot)$ and $\mathcal C(\cdot)$ stand respectively for the {\it cumulants/moments} transform and the {\it moments/cumulants} transform that both take as argument a vector of size $k$ (the first $k$ moments or cumulants, respectively) and output a size $k$ vector (the first $k$ cumulants or moments, respectively).
The inner $\frac{1}{c}\mathcal C(c\cdot)$ operation multiplicatively deconvolve the Marchenko-Pastur law $\mu_{\eta_c}$. Then the next $\mathcal C$ application turns the resulting moments into cumulants. The additive deconvolution of $\delta_{\sigma^2}$ is then performed through the cumulant difference and the output is turned back into the moment space through the $\mathcal M$ application. The outer $\frac{1}{c}\mathcal M(c\cdot)$ is finally performed to multiplicatively convolve the result with $\mu_{\eta_c}$.
\item the moments $m''_k$ of $\mu_{\frac{1}{L}{{\bf P}^{\frac{1}{2}}{\bf H}^{\sf H}{\bf H}{\bf P}^{\frac{1}{2}}}{\bf \Theta\Theta}^{\sf H}}$ are then given through equation \eqref{eq:zeros} by a simple scaling of the moments $m'_k$ by $1/M$. This reads
\begin{align}
(m_1'',\ldots,m_M'')&=\frac{1}{M}(m_1',\ldots,m_M')
\end{align}
\item the Marchenko-Pastur law $\mu_{\eta_{c'}}$, $c'=MN/L$, is then deconvolved from $\mu{\frac{1}{L}{{\bf P}^{\frac{1}{2}}{\bf H}^{\sf H}{\bf H}{\bf P}^{\frac{1}{2}}}{\bf \Theta\Theta}^{\sf H}}$ according to equation \eqref{eq:lastdec}. This leads then to the moments $m_k'''$ of $\mu_{{\bf P}^{\frac{1}{2}}{\bf H}^{\sf H}{\bf H}{\bf P}^{\frac{1}{2}}}$,
\begin{align}
(m_1''',\ldots,m_M''')&=\frac{1}{c'} \mathcal C\left( c'd_1',\ldots,c'd_M' \right) 
\end{align}
\item finally, the resulting moments $m_k'''$ are scaled by $M$ to obtain the $d_k$ coefficients,
\begin{align}
(d_1,\ldots,d_M)&=M(m_1''',\ldots,m_M''')
\end{align}
\end{enumerate}

\begin{figure}
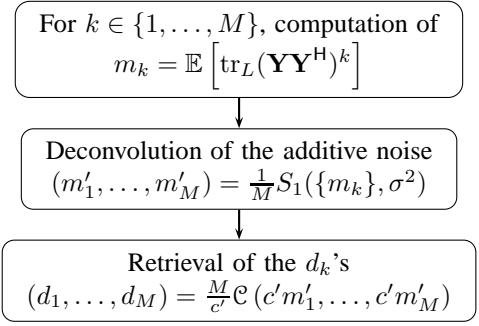

\centering
\begin{psmatrix}[mnode=r,colsep=0.8,rowsep=0.4]
[name=e0] \pw{\begin{tabular}{c}For $k\in \{1,\ldots,M\}$, computation of \\ $m_k=\mathbb E\left[\tr{L}({\bf YY}^{\sf H})^k\right]$ \end{tabular}} \\ [0pt]
[name=e1] \pw{\begin{tabular}{c}Deconvolution of the additive noise \\ $(m_1',\ldots,m_M')=\frac{1}{M}S_1(\{m_k\},\sigma^2)$ \end{tabular}} \\ [0pt]
[name=e2] \pw{\begin{tabular}{c} Retrieval of the $d_k$'s \\ $(d_1,\ldots,d_M)=\frac{M}{c'} \mathcal C\left( c'm_1',\ldots,c'm_M' \right)$ \end{tabular}} \\ [0pt]

\ncline{->}{e0}{e1}
\ncline{->}{e1}{e2}
\end{psmatrix}
\caption{$m_k$ to $d_k$ Block Diagram}
\label{fig:bdMD}
\end{figure}

Figure \ref{fig:bdMD} summarizes these steps.

Our final interest though is to find the diagonal values of $\bf P$. The distribution of the channel matrix ${\bf H}$ was modelled in Section \ref{sec:model} by a mixture of correlated Gaussian subchannels. It is difficult, at this point of knowledge in free probability theory, to deconvolve the effect of ${\bf H}$ from the random matrix ${\bf H}{\bf P}{\bf H}^{\sf H}$. Only classical methods can help in this situation. Remarkably, it turns out that the matrix ${\bf H}{\bf P}{\bf H}^{\sf H}$ is diagonal
\begin{equation}
{\bf H}{\bf P}{\bf H}^{\sf H} = \begin{bmatrix} 
\sum_{k} P_k|h_{k1}|^2  &  \cdots          & 0 \\
\vdots                             & \ddots & \vdots \\
0                                  & \cdots & \sum_{k} P_k|h_{kN}|^2 \\
\end{bmatrix}
\end{equation}

Therefore the theoretical moments $d_k$ of $\mu_{{\bf H}{\bf P}{\bf H}^{\sf H}}$ are the normalized traces of the asymptotic diagonal matrices of entries,
\begin{equation}
\left\{({{\bf H}{\bf P}{\bf H}^{\sf H})}^k\right\}_{ij} = \left(\sum_{k=1}^M P_k|h_{ki}|^2\right)^k \delta_i^j
\end{equation}
and then the $p^{th}$ order moment $d_p=\EE[\tr{N}{({\bf HPH}^{\sf H})^p}]$ of ${\bf H}{\bf P}{\bf H}^{\sf H}$ can then be equated for large $N$ to
\begin{equation}
\label{eq:dp1}
d_p = \lim_{N\to \infty} \frac{1}{N}\sum_{j=1}^N \left(\sum_{k=1}^M P_k|h_{kj}|^2\right)^p
\end{equation}

In the following we provide a method to estimate the entries $P_k$ under the assumption, which is never verified in practice, that the channels are extremely frequency selective, i.e. such that, for any $k$ and any couple $j\neq j'$, the entries $h_{k,j}$ and $h_{k,j'}$ are independent.

\subsection{Estimation of the powers $P_k$}
\label{sec:findPk}

As already concluded in Section \ref{sec:info}, if the channel delay spreads are very short, then the channel frequency responses $\{h_{k1},\ldots,h_{kN}\}$, $k\in \{1,\ldots,M\}$, are strongly correlated and almost nothing can be deduced on the entries of ${\bf P}$. On the contrary, assume that the channel delay spread is very large, and rewrite equation \eqref{eq:dp1} as
\begin{align}
\label{eq:dp2}
d_p &= \bar{d}_p + w_p \\
&=\frac{1}{N}\sum_{j=1}^N \left(\sum_{k=1}^M P_k|h_{kj}|^2\right)^p + w_p
\end{align}
with $\bar{d}_p$ the $p^{\rm th}$ sample order moment ($N<\infty$) and $w_p$ some noise process. The latter converges to $0$ when $N\to \infty$ and the channel frequency response is i.i.d. Gaussian. However, when $N$ is finite or when the channel is less frequency selective, then $w_p$ is a bit more difficult to handle.

To push the computation forward, we need first to derive the classical order $p$ moment $m^{(h)}_p$ of the variables $|h_{kj}|^2$, for any couple $(k,j)$, in the complex case. This gives
\begin{equation}
\label{eq:mom}
m^{(h)}_p = \EE[|h_{kj}|^{2p}] = \frac{1}{2^{2p}}\sum_{i=0}^pC_p^i\frac{(2i)!(2[p-i])!}{i!(p-1)!}
\end{equation}

One can then derive the general expression for $d_p$ as
\begin{align}
\label{eq:dp}
d_p &= \frac{p!}{2^{2p}}\sum_{\substack{k_1,\ldots,k_M \\ \sum_i k_i=p}}\prod_{i=1}^M\left\{ \sum_{k=0}^{k_i}\frac{(2k)!(2[k_i-k])!}{(k!)^2([k_i-k]!)^2}\right\}P_i^{k_i} 
\end{align}
The complete derivation of formula \eqref{eq:dp} is provided in the appendix.

\subsubsection{Bayesian approach}
\label{sec:bayes}
Let $K$ be some integer greater than or equal to $M$. The noise process ${\bf w}=[w_1,\ldots,w_K]^{\sf T}$, as already mentioned, is in general difficult to analyze. We shall consider in the following that one actually has a limited knowledge about ${\bf w}$ which reduces to the covariance matrix ${\bf C}={\rm E}[{\bf w}{\bf w}^{\sf T}]$ gathered from previous simulations\footnote{honesty would require that we actually derive the maximum entropy distribution of $\bf w$ but this would lead to involved computation.}. Now, consider that the set of prior information $I$ contains the following elements:
\begin{itemize}
  \item the values of $P_1,\ldots,P_M$ are real positive and known not to be larger than some value $P_{\sf max}$.
  \item the typical error variance ${\bf C}$ in the observed moments $\{d_k\}$, $k\in \{1,\ldots,M\}$, is known.
\end{itemize}

In fact, the exact covariance matrix ${\bf C}$ cannot be known since its computation requires the exact knowledge of ${\bf P}$. Indeed, a few lines of calculus of 
\begin{equation}
  {\bf C} = {\rm E}\left[\left( {\bf d} - \bar{\bf d} \right)^{\sf H}\left( {\bf d} - \bar{\bf d} \right)\right]
\end{equation}
with $\bar{\bf d}=[\bar{d}_1,\ldots,\bar{d}_K]^{\sf T}$, lead to equation \eqref{eq:C}, which depends on $P_1,\ldots,P_M$.

\begin{figure*}[t!]
\begin{align}
  \label{eq:C}
  C_{a,b} &=-\bar{d}_a\bar{d}_b + \sum_{\substack{k_1,\ldots,k_M \\ k_1',\ldots,k_M' \\ \sum_i k_i=a \\ \sum_j k_j=b}} \left\{ \frac{(N-1)a!b!}{N} \prod_{\substack{i,j \\ 1\leq i \leq a \\ 1\leq j\leq b}} \frac{P_{k_i}^{k_i}P_{k_j'}^{k_j'}m^{(h)}_{k_i}m^{(h)}_{k_j'}}{k_i!k_j'!} + \frac{a!b!}{N}\prod_{\substack{i,j \\ 1\leq i \leq a \\ 1\leq j\leq b\\ k_i=k_j'}} \frac{P_{k_i}^{2k_i}m^{(h)}_{2k_i}}{(k_i!)^2} \times \prod_{\substack{i,j \\ 1\leq i \leq a \\ 1\leq j\leq b \\ k_i\neq k_j'}} \frac{P_{k_i}^{k_i}P_{k_j'}^{k_j'}m^{(h)}_{k_i}m^{(h)}_{k_i}}{k_i!k_j'!}\right\}
\end{align}
\hrulefill
\end{figure*}

However, let us first consider ${\bf C}$ known before we introduce alternative solutions when $\bf C$ is unknown.

The objective is to infer on the set $\{P_1,\ldots,P_M\}$ given $I$ and the observed sample moments $\bar{\bf d}$. An error measure must be considered to come up with an estimate of $\{P_1,\ldots,P_M\}$. We consider here the estimate of $\bf P$ which minimizes the mean quadratic error (MMSE).
This MMSE estimate $\tilde{\bf P}$ is given by
\begin{align}
  \tilde{\bf P} &= {\rm E}\left[ {\bf P}|\bar{\bf d} \right] \\
  &= \int_{\bf P} {\bf P} p({\bf P}|\bar{\bf d},I) {\rm d}{\bf P} \\
  \label{eq:Pbayes}  &= \int_{\bf P} {\bf P} \frac{p(\bar{\bf d}|{\bf P},I) p({\bf P}|I)}{\int_{\bf P} p(\bar{\bf d}|{\bf P},I)p({\bf P}|I) {\rm d}{\bf P}} {\rm d}{\bf P}
\end{align}

Since the prior information $({\bf P}|I)$ is limited to the fact that all entries are upper-bounded by $P_{\sf max}$, $p({\bf P}|I)$ should be set uniform on the space $[0,P_{\sf max}]^M$ according to the maximum entropy principle. However, note that if $\{\tilde{P}_1,\ldots,\tilde{P}_M\}$ minimizes the MMSE then also does any permutation of this set. Therefore, to have a correctly defined problem (with a unique solution), the set $\{P_1,\ldots,P_M\}$ must be ordered; we will then state in the following that $P_1\leq P_2\leq P_M$. Therefore the prior $p(P_k|I)$ is taken uniformly on the set $[0,P_{k-1}]$ when $P_{k-1}$ is set, which leads to $p({\bf P}|I)=P_{\sf max}\prod_{i=1}^{M-1} P_i^{-1}$. Also, since only the error covariance matrix ${\bf C}$ in the observed sample moments $\bar{\bf d}$ is known, the maximum entropy principle requests that the process ${\bf w}$ is assigned a Gaussian distribution with variance $\bf C$. Therefore, equation \eqref{eq:Pbayes} becomes
\begin{align}
  \label{eq:Pbayesfinal}
  \tilde{\bf P} &= \int_{P_1\leq\ldots \leq P_M} {\bf P} \frac{e^{-{\bf w}({\bf P})^{\sf H}{\bf C}^{-1}{\bf w}({\bf P})}}{\int_{P_1\leq\ldots \leq P_M} e^{-{\bf w}({\bf P})^{\sf H}{\bf C}^{-1}{\bf w}({\bf P})} {\rm d}{\bf P}} {\rm d}{\bf P}
\end{align}
where we denoted ${\bf w}={\bf w}({\bf P})$ to remind the actual dependence of ${\bf w}$ in the powers $P_1,\ldots,P_M$ (through the expression of ${\bf d}$ in equation \eqref{eq:dp}).

Unfortunately, the integration space of equation \eqref{eq:Pbayesfinal} makes both integrals rather involved to compute. A way to practically computed $\tilde{\bf P}$ consists in turning the integrals into finite sums over thin sliced versions of the integration space.
Also, as previously mentioned, the covariance matrix $\bf C$ is obviously unknown when trying to decipher the cell powers $P_1,\ldots,P_M$. However, iterative methods can be considered in which $\bf C$ is initially defined as the covariance matrix ${\bf C}_{\sf init}$ of an hypothetic set of powers, say $P_1=P_2=\ldots =P_M=P_{\sf max}/2$. Then, running the MMSE estimator with ${\bf C}_{\sf init}$ returns a first set $P^{(1)}_1,\ldots,P^{(1)}_M$ from which a refined version of $\bf C$ can be evaluated (from formula \eqref{eq:C}). Note that, in running $k$ instances of the algorithm, the sample moments $\tilde{d}_p$ can be accumulated and the covariance matrix $\bf C$ has to be computed as if as many as $kN$ subcarriers were actually used in the free deconvolution algorithm. This process can be processed in a loop for a satisfying number of iterations.

\subsubsection{Alternative estimators}
Other estimators than MMSE, such as maximum-likelihood (ML), might be considered which take as an estimate the set ${\bf P}$ which minimizes ${\bf w}({\bf P})^{\sf H}{\bf C}^{-1}{\bf w}({\bf P})$. However, the measure associated to the ML estimator does not suit the broad {\it a posteriori} distribution $p({\bf P}|\bar{\bf d},I)$ as will be shown in simulations. Indeed, a large estimation error is as bad as a small estimation error in the ML context; therefore, when the posterior $p({\bf P}|{\bf d},I)$ is not peaky, large estimation errors are expected. The MMSE estimator is, in this scenario, more appropriate.

A zero-forcing method can also be derived. From equation \eqref{eq:dp}, if nothing were known about the noise process $w_p$, one might naively consider solving the system of $M$ equations \eqref{eq:dp}, with $p=1,\ldots,M$ in the $M$ unknowns $P_1,\ldots,P_M$, in which $d_p$ is set equal to the observed $\tilde{d}_p$. This can be solved by turning this system of equations into the equivalent system
\begin{align}
\label{eq:systemeq}
\sum_{k=1}^{M} P_k   &= Q_1(d_1) \nonumber \\
\sum_{k=1}^{M} P_k^2 &= Q_2(d_1,d_2) \nonumber \\
\vdots~~~~~~ &= ~~~~~~~~\vdots \nonumber \\
\sum_{k=1}^{M} P_k^M &= Q_{M}(d_1,\ldots,d_{M})
\end{align}
with $Q_k\in R[d_1,\ldots,d_k]$.

This system can be solved using the Newton-Girard formulas \cite{SER00}; the solutions $P_1,\ldots,P_M$ are found to be the $M$ roots of an $M^{\rm th}$ degree polynomial. This solution does not require any knowledge on the covariance matrix $\bf C$, which does not have any signification in this context. However, it often turns out that the roots $P_k$ are not all real, which makes the solution useless in practice. Also, this zero-forcing method is very awkward as it strongly suffers from the presence of noise. Especially, the large variance ${\rm E}[|w_M|^2]$ is considered equally to the typically very small variance ${\rm E}[|w_1|^2]$, and therefore the first sample moment $\bar{d}_1$ is not more important in the final computation than the last sample moment $\bar{d}_M$\footnote{this would be here again a dishonest consideration as one at least knows that $d_M$ is more uncertain than $d_1$.}.

\section{Simulation and Results}
\label{sec:results}
In the following, we use the results that were previously derived in the case of a three-cell network that the terminal wishes to track. The set of cells studied along this part are of relative powers $P_1=4$, $P_2=2$, $P_3=1$. 


\begin{figure}
  \begin{center}
	  \includegraphics[width=8cm]{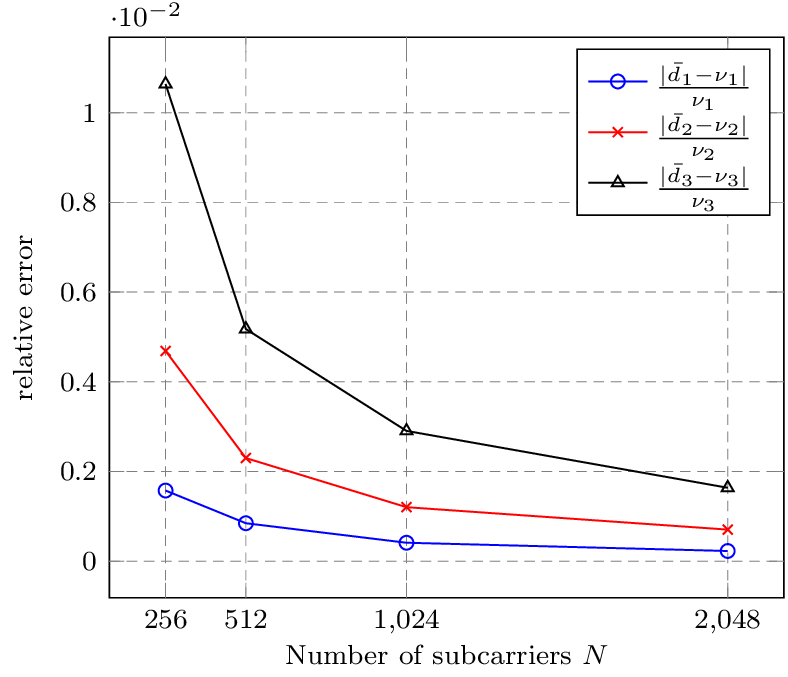}
  \caption{Relative error on recovered moments $\nu_k=\tr{N}({\bf HPH}^{\sf H})^k$ from the free deconvolution algorithm, $L=2N$}
\label{fig:relerror}
\end{center}
\end{figure}

\begin{figure}
  \begin{center}
	  \includegraphics[width=8cm]{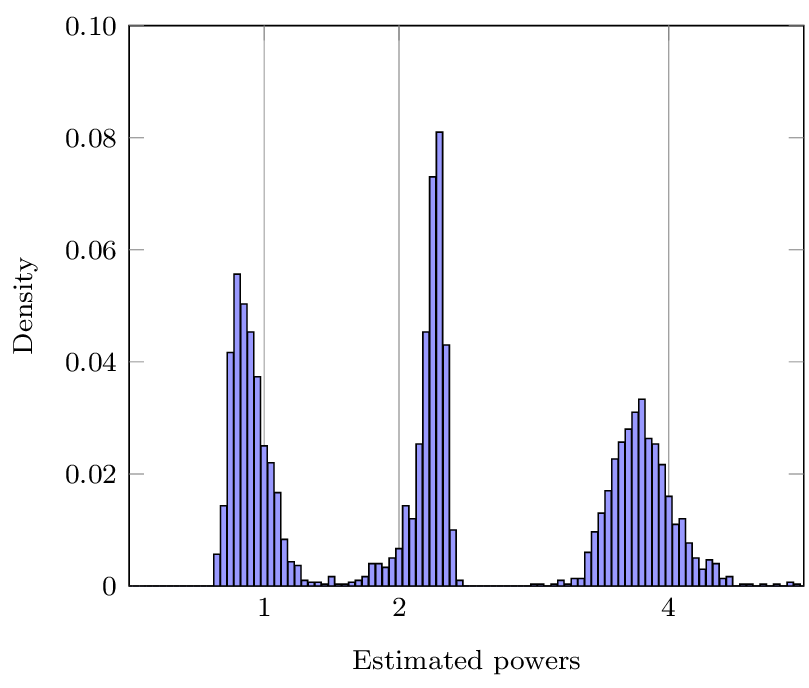}
  \caption{Cell power detection, $N=2048$, $L=4096$, MMSE estimate, Perfect knowledge of $\bf C$}
  \label{fig:2048_noaccum}
\end{center}
\end{figure}

\begin{figure}
  \begin{center}
  \begin{tabular}{cc}
	  \includegraphics[width=4cm]{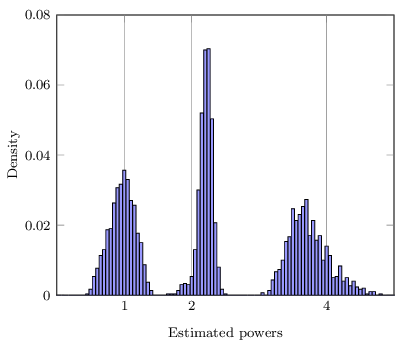}
&
	  \includegraphics[width=4cm]{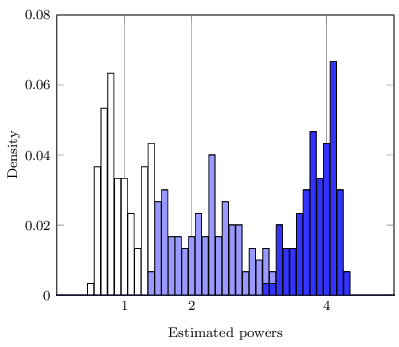}
\end{tabular}
\caption{Cell power detection, $N=512$, $L=1024$, MMSE estimate (left) and ML estimate (right), Perfect knowledge of $\bf C$}
  \label{fig:512_noaccum_ML}
\end{center}
\end{figure}

\begin{figure}
  \begin{center}
	  \includegraphics[width=8cm]{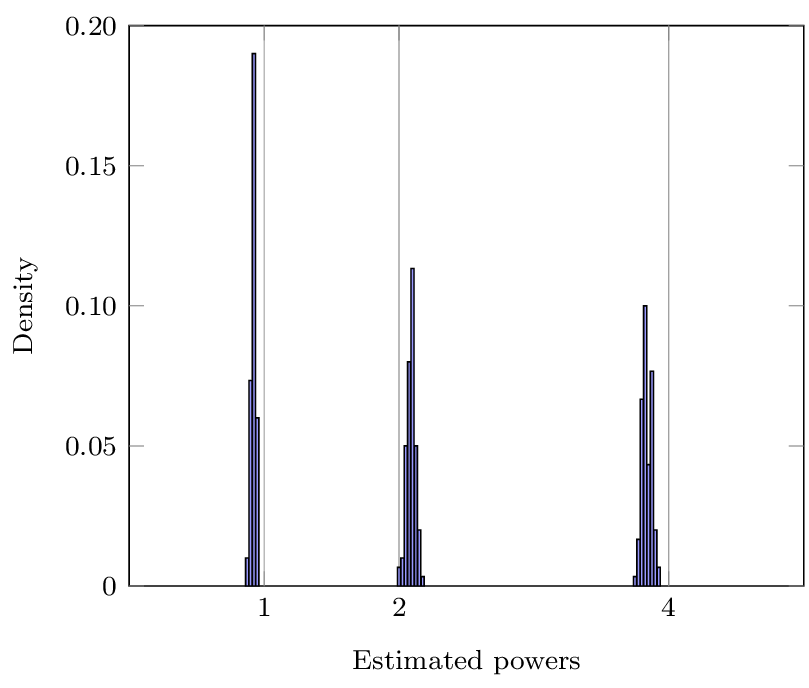}
  \caption{Cell power detection, $N=512$, $L=1024$, $10$ accumulations, MMSE estimate, Perfect knowledge of $\bf C$}
  \label{fig:512}
\end{center}
\end{figure}
\begin{figure}
  \begin{center}
	  \includegraphics[width=8cm]{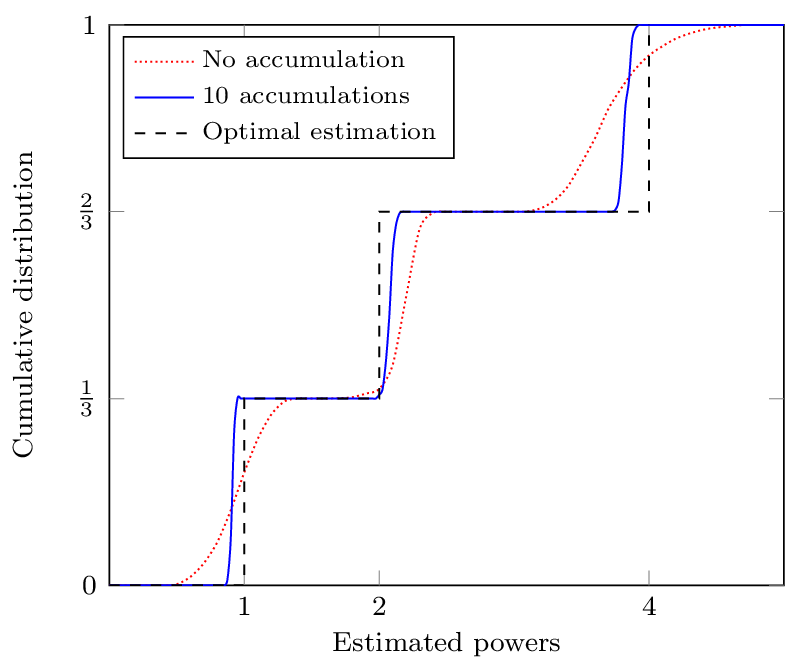}
  \caption{Cell power detection CDF, $N=512$, $L=1024$, Perfect knowledge of ${\bf C}$}
\label{fig:512cdf}
\end{center}
\end{figure}

Before performing the first simulation of the complete algorithm, we present in Figure \ref{fig:relerror} the relative error in the estimation of the moments $\nu_k=\tr{N}({\bf HPH}^H)^k$ from the free deconvolved sample moments $\bar{d}_k$. It is observed that even for $N=256$, $L=512$ the mean relative error in the computed third moments is of order $1\%$. This suggests that the free deconvolution technique is very accurate even for non-infinite values of $N$ and $L$. The bottleneck approximation in the estimation of $P_1,\ldots,P_M$, as will be observed in the coming plots, therefore lies in the convergence of the entries $h_{kj}$ towards a Gaussian i.i.d. process, and not in the infinite matrix size assumption.

In a first simulation, we study the convergence properties of the proposed scheme. We consider a large OFDM system with $N=2048$ subcarriers and $L=4096$ sampling periods under ideal Gaussian i.i.d. input symbols, uncorrelated Gaussian channel frequency responses and Gaussian additive white noise with signal to noise ratio ${\sf SNR}=20~{\rm dB}$. The covariance matrix $\bf C$ is the exact covariance matrix. Figure \ref{fig:2048_noaccum} provides the results of this simulation for thousand channel realisations. It is observed that the distribution of the eigenvalues is largely spread over the expected eigenvalues. This is explained by the slow convergence nature of the computed $d_p$ towards the corresponding moments when $N\to \infty$. Also, it is observed that the peak centers are offset from the expected powers. This is mainly due to the fact that the MMSE estimator is meant to minimize the {\it mean} quadratic error averaged over all possible sets $P_1,P_2,P_3$ and not for the particular set selected here\footnote{it was observed in particular from other simulations that cells of equal powers are generally better estimated.}. Also, the actually non-Gaussian property of the noise process $\bf w$ as well as the inexact results from the free deconvolution process contribute to the offset. 

Figure \ref{fig:512_noaccum_ML} provides a comparison between the results given by the MMSE estimator and the ML estimator when $N=512$ and $L=1024$. It turns out, as discussed earlier, that the ML estimate is more largely spread around the expected cell powers than the MMSE estimate.

In the following, we perform more realistic simulations in which input signals are QPSK modulated instead of complex Gaussian, and with more realistic channels of length varying from $1$ to $N/4$ symbols. To reduce the variance of the estimates, we also average the sample $d_p$ over ten channel realizations, which in practice requires around $1~{\rm ms}$ of data to process. In Figure \ref{fig:512}, we took $N=512$, $L=1024$ and a Rayleigh channel of length $N/8$. The covariance matrix $\bf C$ is still the exact covariance matrix. Then a hundred realisations of this process are run. The SNR is still ${\sf SNR}=20~{\rm dB}$ in this second experiment. The cumulative distribution function (CDF) of the detected powers is presented in Figure \ref{fig:512cdf} and compared to the CDF when no accumulation is performed. The three thresholds corresponding to the three detected cell powers can be observed, with a slight shift from the expected cell powers caused both by the non-exact Gaussian assumption on $\bf w$ and on the non-exact Gaussian i.i.d. assumption on the channel frequency responses.

\begin{figure}
  \begin{center}
	  \includegraphics[width=8cm]{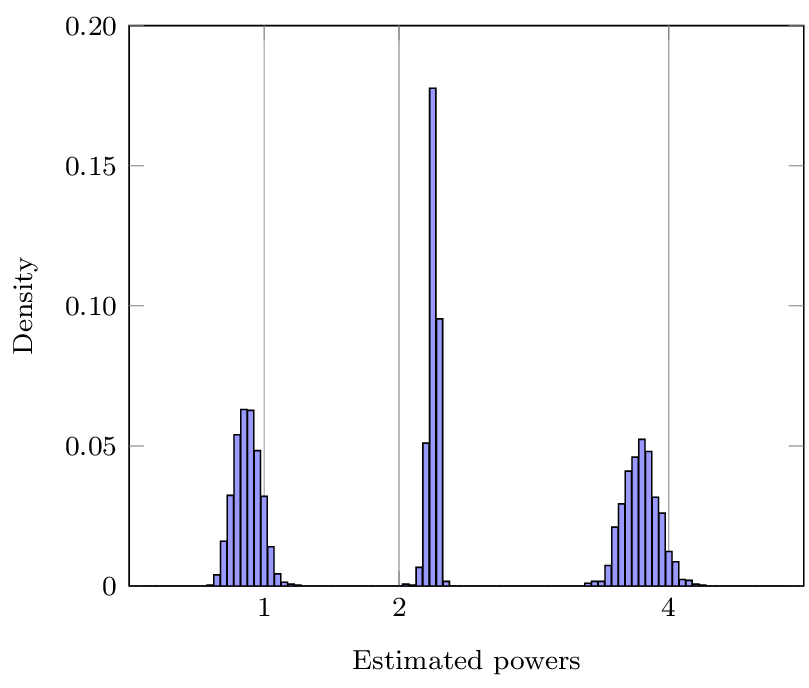}
  \caption{Cell power detection, $N=512$, $L=1024$, MMSE estimate, Recursive update of $\bf C$, $10$ steps}
  \label{fig:iter}
  \end{center}
\end{figure}

Now we propose to examine the performance of the iterative cell power recovery. We initially set ${\bf C}_{\sf init}$ to the covariance matrix of a set of cells of powers $P_1=P_2=P_3=2.5$. Then ten iterations of the cell-power detector are run, with at each step a refinement of $\bf C$. Note that, since at each step the sample moments $\bar{d}_p$ are accumulated, the entries of the covariance matrix $\bf C$ are computed accordingly, i.e. $k$ accumulations demand that $\bf C$ is computed from an effective number $kN$ subcarriers, as presented in Section \ref{sec:algo}. Figure \ref{fig:iter} provides the results of the recursive algorithm, which proves to performs very accurately, with surprisingly little dispersion around the detected powers compared to Figure \ref{fig:512}.

\begin{figure}
  \begin{center}
	  \includegraphics[width=8cm]{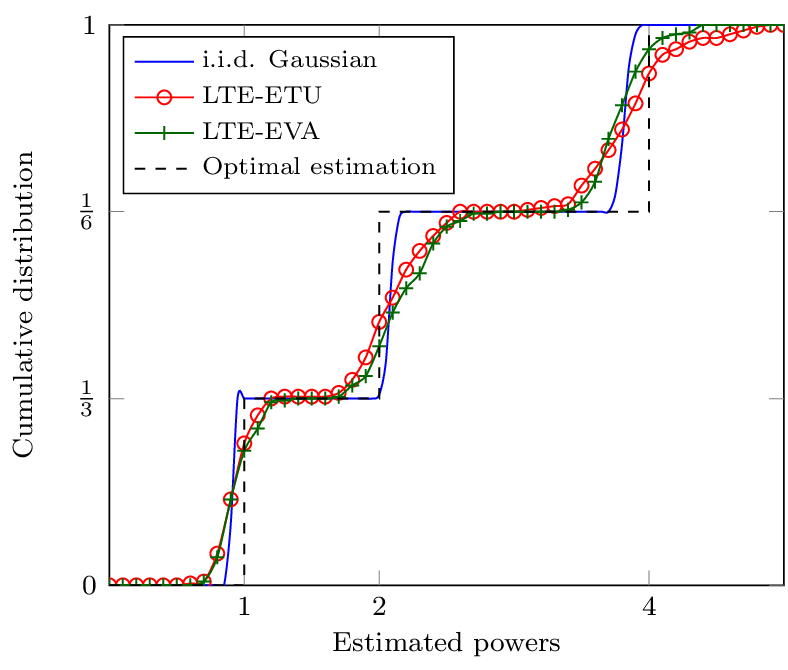}
  \caption{Cell power detection CDF in LTE channels, $N=512$, $L=1024$, $10$ accumulations, preset $\bf C$}
\label{fig:LTECh}
  \end{center}
\end{figure}

Also, we test the robustness of our algorithm against practical short channels, instead of high delay spread channels. This is shown in Figure \ref{fig:LTECh} which proposes a comparison between the ideal long channel situation and the 3GPP-LTE \cite{LTE} standardized Extended Vehicular A (EVA) and Extended Typical Urban (ETU) channels with parameters given in Table \ref{tab:LTE}.

\begin{table}
\centerline{
\begin{tabular}{ccc}
\toprule
{Channel Type} & {RMS delay spread} & {Channel length} \\ 
\midrule
EVA & $357~{\rm ns}$ & $N/27$ \\
ETU & $991~{\rm ns}$ & $N/13$ \\
\bottomrule
\addlinespace[2\aboverulesep]
\end{tabular}
}
\caption{3GPP-LTE standardized short delay channels}
\label{tab:LTE}
\end{table}

Here we considered a mobile handset with $N_r=2$ antennas, $256$ subcarriers per antenna (and then in total an effective number of $N=512$ subcarriers), $L=1024$, ${\sf SNR}=20~{\rm dB}$. The covariance matrix $\bf C$ of the noise process $\bf w$ is an approximated matrix obtained from intensive simulations on short channels. Indeed, the covariance pattern is very different from the matrices used in the Gaussian i.i.d. scenario and is difficult to derive analytically; especially we noticed that the smaller the delay spread, the larger the uncertainty on the higher moments, which turns $\bf C$ into an ill-conditioned matrix. The results are averaged over ten channel realizations. The CDF of the detected power distribution for those channels is provided in Figure \ref{fig:LTECh}. The latter shows a rather good behaviour in both ETU and EVA channels. Nonetheless their short delay spreads lead to a larger variance in the mean power estimation.

Surprisingly, it turns out that the distribution of the input signals ${\bf s}^{(l)}$ does not impact the system performance as long as it is i.i.d. with zero mean and unit variance. This is a known result in free deconvolution which has not been proven yet. Therefore in our simulations, QPSK modulations showed the exact same behaviour as purely Gaussian distributed input signals. 

\section{Discussion}
\label{sec:discussion}
\subsection{Data, prior and convergence}
We previously described and simulated a recursive algorithm meant to converge to an accurate estimate of the cell powers $P_1,\ldots,P_M$, whose convergence we did not prove yet. Actually, a mathematical as well as a philosophical reasons for this convergence can be advanced. First, note that a large number of iterations lead to a smaller variance of ${\bf w}={\bf d}-\hat{\bf d}({\bf P})$, therefore to smaller entries of the noise covariance matrix ${\bf C}$. As a consequence ${\bf C}^{-1}$ has large entries and the exponential terms $e^{-\frac{1}{2}({\bf d}-\hat{\bf d}({\bf P}))^{\sf T}{\bf C}^{-1}({\bf d}-\hat{\bf d}({\bf P}))}$ in the MMSE integrals \eqref{eq:Pbayesfinal} are relevant only when the differences ${\bf d}-\hat{\bf d}({\bf P})$ are very small; this is, when $\hat{\bf d}({\bf P})\simeq {\bf d}$. Therefore the accuracy in the entries of ${\bf C}^{-1}$ is not of fundamental importance, as long as they are large enough. The convergence of the iterative process is ensured by the accumulated data ${\bf d}$ themselves; but of course this convergence is accelerated with good approximations of ${\bf C}$.

This observation is very general in the Bayesian probability theory context and has been observed and analyzed thoroughly by Jaynes \cite{JAY03}. Bayesian probabilities rely on a balance between {\it priors} and {\it data}. If the available data is scarce, then prior information is very valuable; in our situation, as shown by simulation, if $\bf C$ were known, then even a single channel realization allows to approximately recover the transmitted powers. On the contrary, large amounts of data prevail over prior information so that even unfortunate priors may not badly alter the {\it a posteriori} probability; this explains why a precise estimation of ${\bf C}$ is not mandatory when large accumulations of data are considered. Note that in this case, ML estimates and MMSE estimates show similar performance, since the posterior distribution $p({\bf P}|{\bf Y},I)$ is very peaky in the correct value for ${\bf P}$.

However, it is important to underline the fact that the offset problem, observed in simulation, in the estimation of $P_1,\ldots,P_M$ will not be reduced by mere data accumulations. Indeed, the issue comes here from the free deconvolution process which is not accurate for finite $N$. Therefore, an appropriate trade-off between large $N$ and many accumulations must be found: large $N$ entail more accurate free deconvolution processes at the expense of computationally demanding large matrix products, while many accumulations ensures a faster convergence (to offset cell powers) at a lower computation cost. Another approach consists in computing higher order moments to strengthen the cell power estimation. However, high order sample moments have a large variance for finite $N$. Their impact on the final estimation might then be very limited since they are very unreliable. For instance we observed in simulations that, for $N=512$, typical matrices $\bf C$ verify $C_{11}\simeq 1e^{-2}$ and $C_{33}\simeq 1e^4$, which gives a million times more credit to the first order sample moment than to the third order sample moment. Bringing in the computation fourth order moments with $N=512$ would turn out not worth the computation increase.

\subsection{Applicability}
As reminded in the introduction of this paper, usual cell power detection techniques use scarce and largely interfered synchronization sequences. Much time, but low computation, is then required to detect cells with high efficiency. From the authors' experience in the context of 3GPP-LTE, many problems arise when those pilot sequences are synchronized and the emitting cells have equal powers since both signals interfere to the detriment of the decoder. In the previously derived scheme, even cells of equal power can be counted and isolated. Indeed, if $M$ cells are expected and the CDF of the estimated powers shows a large jump of $2/M$ for a given power $P$, then we can deduce that two cells have equal power $P$. It is therefore not necessary to enhance the quality of the estimation to {\it separate} those two cells.

However, some limitations can be found in the applicability of this cognitive scheme. Firstly, loaded cells are required to ensure reliability of the estimates; this requires that the underlying standards optimally reuse the allocated bandwidth. Secondly, if many cells are desired to be detected, then very high order moments must be computed which, as discussed earlier are only reliable if the number of subcarriers $N$ and the number of accumulations are very large. High order moments also require very large matrix products, which might be too demanding to the embedded system processors. Also, under the limited state of knowledge of the system model, neighboring cells can only be detected if the propagation channels are very frequency selective. The flatter the channels, the more numerous the accumulations required to come up with a reliable estimation. The latter limitation is obviously the most constraining factor.

\section{Conclusion}
\label{sec:conclusion}
In this paper, we provided a practical way to blindly detect neighboring cells in a distributed OFDM network. Assuming constant transmission in those cells on a fairly large bandwidth, we showed that one can determine the individual SNR of every surrounding cell provided that the channel delay spread is sufficiently large. This scheme is particularly suited for the future cognitive OFDM systems which aim to reduce the amount of synchronization sequences while keeping track of the neighboring cells.

\section{Acknowledgement}
This work was partially supported by the European Commission in
the framework of the FP7 Network of Excellence in Wireless
Communications NEWCOM++.

\appendix
\section{Proof of formula \eqref{eq:dp}}
\label{app:dp}
Consider Gaussian channels with independent frequency responses $h_{ij}$ ($i\in \{1,\ldots,M\}$, $j\in [1,N]$). Then, for a given $j$, noise taken apart, we denote 
\begin{align}
d^{(j)}_p &=\left( \sum_{i=1}^M P_i|h_{ij}|^2\right)^p \nonumber \\
&= \sum_{\substack{k_1,k_2,\ldots,k_M \\ \sum_m k_m=p}}C_p^{k_1}C_{p-k_1}^{k_2}\ldots C_{p-k_1-\ldots -k_{M-1}}^{k_M}\prod_{l=1}^M(P_i|h_{ij}|^2)^{k_l}
\end{align}
where the product of the binomial coefficients $C_p^{k_1}C_{p-k_1}^{k_2}\ldots C_{p-k_1-\ldots -k_{M-1}}^{k_M}$ is the multinomial coefficient 
$\frac{p!}{k_1!k_2!\ldots k_M!}$.

Hence the simplified expression, when averaging $d^{(j)}_p$ over $j\in [1,N]$ (with $N\to \infty$ to ensure $\frac{1}{N}\sum_j |h_{ij}|^2 \to \mathbb E_j[|h_{ij}|^2]$)
\begin{align}
d_p &= p! \sum_{\substack{k_1,k_2,\ldots,k_M \\ \sum_m k_m=p}} \frac{\mathbb E_j\left[\prod_{i=1}^M (P_i|h_{ij}|^2)^{k_i} \right] }{\prod_{i=1}^M k_i!}
\end{align}

The moments of the complex variable $|h_{ij}|^2$ are deduced from the moments of a Gaussian real variable by considering
\begin{equation}
|h_{ij}|^2 = (h_{ij}^{\Re}+ i\cdot h_{ij}^{\Im})(h_{ij}^{\Re}- i\cdot h_{ij}^{\Im})
\end{equation}
where the real and imaginary parts $h_{ij}^{\Re}$ and $h_{ij}^{\Im}$ of $h_{ij}$ are standard Gaussian variables whose even moments are known
\begin{equation}
\mathbb E[(h_{ij}^{\Re})^{2p}] = \mathbb E[(h_{ij}^{\Im})^{2p}] = \frac{(2p)!}{2^p p!}
\end{equation}
hence equation (\ref{eq:mom}).

This results in
\begin{equation}
d_p = \frac{p!}{2^{2p}}\sum_{\substack{k_1,\ldots,k_M \\ \sum_i k_i=p}}\prod_{i=1}^M\left\{ \sum_{k=0}^{k_i}\frac{(2k)!(2[k_i-k])!}{(k!)^2([k_i-k]!)^2}\right\}P_i^{k_i}
\end{equation}


\begin{thebibliography}{1}
\bibitem{SHA48} C. E. Shannon, ``A mathematical theory of communications'', Bell System Technical Journal, vol. 27, no. 7, pp. 379-423, 1948.
\bibitem{PRO98} E. Biglieri, J. Proakis, and S. Shamai, ``Fading channels: Information-theoretic and communications aspects,'' IEEE Trans. on Inf. Theory, vol. 44, no. 6, pp. 2619-2692, Oct 1998.
\bibitem{WIN98} J. H. Winters, ``Smart Antennas for Wireless Systems'', IEEE Personnal Communications, vol. 5, no. 1, pp.23-27, Feb. 1998.
\bibitem{LTE} http://www.3gpp.org/Highlights/LTE/LTE.htm
\bibitem{WiMax} http://wirelessman.org/
\bibitem{KAR06} K. Karakayali, G. J. Foschini, R. A. Valenzuela, ``Network Coordination for Spectrally Efficient Communications in Cellular Systems'', August 2006, IEEE Wireless Communications Magazine (invited).
\bibitem{RYA07} {\O}. Ryan, M. Debbah, ``Free deconvolution for signal processing applications'', under review, http://arxiv.org/abs/cs.IT/0701025 
\bibitem{RMT} A. M. Tulino, S. Verd\'u, ``Random Matrix Theory and Wireless Communications'', Now Publishers, vol. 1, Issue 1, 2004.
\bibitem{BIA03} P. Biane, ``Free Probability for Probabilists'', Quantum Probability Communications, World Scientist, 2003.
\bibitem{URK67} H. Urkowitz, ``Energy detection of unknown deterministic signals,'' Proc. of the IEEE, vol. 55, no. 4, pp. 523-531, Apr. 1967
\bibitem{KOS02} V. I. Kostylev, ``Energy detection of a signal with Random Amplitude'', Proc IEEE Int. Conf. on Communications (ICC'02). New York City, pp. 1606-1610, May 2002.
\bibitem{HAY05} S. Haykin, ``Cognitive Radio: Brain-Empowered Wireless Communications'', IEEE Journal on Selected areas in Comm. , vol 23., no. 2, Feb 2005.
\bibitem{CAI03} G. Caire and S. Shamai, ``On the achievable throughput of a multiantenna Gaussian broadcast channel'', IEEE Trans. on Information Theory, vol. 49, no. 7, pp. 1691-1706, 2003.
\bibitem{JAY57} E. T. Jaynes, ``Information Theory and Statistical Mechanics'', Physical Review, APS, vol. 106, no. 4, pp. 620-630, 1957.
\bibitem{JAY03} E. T. Jaynes, ``Probability Theory: The Logic of Science'', Cambridge University Press, 2003.
\bibitem{RYA06} {\O}. Ryan, M. Debbah, ``Multiplicative free convolution and information-plus-noise type matrices'', \textit{Submitted to Journal Of Functional Analysis}, 2008, http://www.ifi.uio.no/\~oyvindry/multfreeconv.pdf
\bibitem{DEB07} {\O}. Ryan, M. Debbah, ``Channel Capacity Estimation using Free Probability Theory'', IEEE Trans. on Signal Processing, vol. 56, no. 11, pp. 5654-5667, Nov. 2008. 
\bibitem{SER00} R. Seroul, D. O'Shea, ``Programming for Mathematicians'', Springer, 2000
\bibitem{HAL98} M. Hall Jr. ``Combinatorial Theory'', John Wiley \& Sons, Inc. New York, NY, USA, 1998
\bibitem{BIN90} J.A. Bingham, ``Multicarrier modulation for data transmission: An idea whose time has come,'' IEEE Comm. Mag., vol. 28, pp. 5-14, May 1990.
\bibitem{TSE02} P. Viswanath, D. Tse and R. Laroia, ``Opportunistic beamforming using dumb antennas'', IEEE International Symposium on Information Theory. IEEE. 2002, pp.449
\bibitem{VOI92} D. V. Voiculescu and K. J. Dykema and A. Nica, ``Free Random Variables'', American Mathematical Society, 1992.
\bibitem{MIT99} J. Mitola, and  Jr GQ Maguire, ``Cognitive radio: making software radios more personal'', Personal Communications, IEEE [see also IEEE Wireless Communications] 6(4), 13-18, 1999.
\bibitem{PAP00} C.B. Papadias, ``Globally convergent blind source separation based on a multiuser kurtosis maximization criterion'', IEEE Trans. on Signal Processing, Dec. 2000, pp. 3508-3519.
\bibitem{VIL07} J. Villares, ``Symbol-rate Second-Order Blind Channel Estimation in Coded Transmissions'', in Proc. ICASSP, Honolulu, Hawaii, April 2007.
\bibitem{HER07} C. Herzet, \textit{et al.}, ``Code-Aided Turbo Synchronization'', IEEE Proceedings, Special issue on turbo techniques, 2007.
\bibitem{SLO98} H. Trigui, D.T.M. Slock, ``Cochannel interference cancellation within the current GSM standard'', IEEE ICUPC'98, Florence, Italy, Oct. 1998.
\bibitem{THO05} R. W. Thomas, L. A. DaSilva, A.B. MacKenzie,``Cognitive networks'', DySPAN 2005
\bibitem{LAS08} E. V. Belmega, S. Lasaulce and M. Debbah, "Decentralized Handovers in cellular networks with cognitive terminals", ISCCSP-2008 Malta, 2008.
\bibitem{LAS08b} E. V. Belmega, S. Lasaulce and M. Debbah, "Power Control in
Distributed Multiple Access Channels with Coordination", WNC3, 2008.
\bibitem{COU08} R. Couillet, M. Debbah, ``Multiple Antenna Cognitive Receivers and Signal Detection'', {\it Planned for submission to IEEE Trans. on Information Theory}.
\end{thebibliography}
\end{document}